\documentclass[prd,superscriptaddress,amsfonts,amssymb,amsmath,showpacs,twocolumn]{revtex4-2}
\usepackage{bm}
\usepackage{amsfonts}
\usepackage{latexsym}
\usepackage{graphicx}
\usepackage{amsmath}
\usepackage{palatino}
\usepackage{mathpazo}
\usepackage{textcomp}
\usepackage{float}
\usepackage{booktabs}
\usepackage{dcolumn}
\usepackage{booktabs}
\usepackage{multirow}
\usepackage{hyperref}
\hypersetup{colorlinks,citecolor=blue}
\usepackage{amsmath}
\usepackage{xcolor}
\usepackage{orcidlink}
\usepackage[caption=false]{subfig}
\usepackage{commath}
\usepackage{tabularx}
\def\jnl@style{\it}
\def\aaref@jnl#1{{\jnl@style#1}}

\def\aaref@jnl#1{{\jnl@style#1}}

\def\aj{\aaref@jnl{AJ}}                   
\def\apj{\aaref@jnl{ApJ}}                 
\def\apjl{\aaref@jnl{ApJ}}                
\def\apjs{\aaref@jnl{ApJS}}               
\def\apss{\aaref@jnl{Ap\&SS}}             
\def\aap{\aaref@jnl{A\&A}}                
\def\aapr{\aaref@jnl{A\&A~Rev.}}          
\def\aaps{\aaref@jnl{A\&AS}}              
\def\mnras{\aaref@jnl{Mon.~Not.~Roy.~Astron.~Soc.}}             
\def\prd{\aaref@jnl{Phys.~Rev.~D}}        
\def\prc{\aaref@jnl{Phys.~Rev.~C}}  
\def\prl{\aaref@jnl{Phys.~Rev.~Lett.}}    
\def\qjras{\aaref@jnl{QJRAS}}             
\def\skytel{\aaref@jnl{S\&T}}             
\def\ssr{\aaref@jnl{Space~Sci.~Rev.}}     
\def\zap{\aaref@jnl{ZAp}}                 
\def\nat{\aaref@jnl{Nature}}              
\def\aplett{\aaref@jnl{Astrophys.~Lett.}} 
\def\apspr{\aaref@jnl{Astrophys.~Space~Phys.~Res.}} 
\def\physrep{\aaref@jnl{Phys.~Rep.}}      
\def\physscr{\aaref@jnl{Phys.~Scr}}       
\def\commat{\aaref@jnl{Comm.~Math.~Phys.}}              
\def\science{\aaref@jnl{Science}}               
\def\cqg{\aaref@jnl{Classical Quant.~Grav.}}            
\def\jpcs{\aaref@jnl{JPCS}}                                     
\def\ijmpd{\aaref@jnl{Int.~J.~Mod.~Phys.~D}}                    
\def\grg{\aaref@jnl{Gen.~Relat.~Gravit.}}               
\def\rpp{\aaref@jnl{Rep.~Prog.~Phys.}}          
\def\npa{\aaref@jnl{Nucl.~Phys.~A}}        
\def\lrr{\aaref@jnl{Living Rev.~Rel.}}                   
\def\jcap{\aaref@jnl{J.~Cosmology Astropart.~Phys.}}    
\def\rmp{\aaref@jnl{Rev.~Mod.~Phys.}}   
\def\epjc{\aaref@jnl{Eur.~Phys.~J.~C}}

\allowdisplaybreaks[1]

\begin{document}

\color{black}       

\title{Cosmic expansion beyond $\Lambda$CDM: Investigating power-law and logarithmic corrections}

\author{M. Koussour\orcidlink{0000-0002-4188-0572}}
\email[Email: ]{pr.mouhssine@gmail.com}
\affiliation{Department of Physics, University of Hassan II Casablanca, Morocco.}

\author{N. S. Kavya\orcidlink{0000-0001-8561-130X}}
\email[Email: ]{kavya.samak.10@gmail.com}
\affiliation{Department of P.G. Studies and Research in Mathematics, Kuvempu University, Shankaraghatta, Shivamogga 577451, Karnataka, INDIA}

\author{V. Venkatesha\orcidlink{0000-0002-2799-2535}}
\email[Email: ]{vensmath@gmail.com}
\affiliation{Department of P.G. Studies and Research in Mathematics, Kuvempu University, Shankaraghatta, Shivamogga 577451, Karnataka, INDIA}

\author{N. Myrzakulov\orcidlink{0000-0001-8691-9939}}
\email[Email: ]{nmyrzakulov@gmail.com}
\affiliation{L. N. Gumilyov Eurasian National University, Astana 010008,
Kazakhstan.}

\date{\today}

\begin{abstract}
The cosmic acceleration observed in the expansion of the Universe has sparked extensive research into the nature of dark energy, which is known to constitute approximately 70\% of the Universe's energy content. In this study, we explore two parametrizations of the Hubble parameter, namely power-law and logarithmic corrections, as alternatives to the standard $\Lambda$CDM model. Using observational data from Cosmic Chronometers (CC), Pantheon+, and the Baryonic Acoustic Oscillations (BAO) datasets, we investigate the dynamics of essential cosmological parameters, including the deceleration parameter, energy density, pressure, and equation of state (EoS) parameter. The $Om(z)$ diagnostic test is employed to classify different dark energy models. Our cosmological models, with the power-law and logarithmic corrections, are found to provide a good fit to the recent observational data and efficiently describe the cosmic expansion scenario.
\end{abstract}

\maketitle

\tableofcontents

\section{Introduction}
\label{sec1}

The field of cosmology underwent a significant paradigm shift with the emergence of observational evidence supporting the accelerating behavior of cosmic expansion, as confirmed by Type Ia supernovae (SNeIa) searches \cite{Riess, Perlmutter}. Over the past two decades, a wealth of observational results, including studies of Baryon Acoustic Oscillations (BAO) \cite{D.J., W.J.}, Cosmic Microwave Background (CMB) \cite{R.R., Z.Y.}, Large Scale Structure (LSS) \cite{T.Koivisto, S.F.}, and the Planck collaborations \cite{Planck2020}, have consistently supported the notion of cosmic acceleration. The prevailing explanation for this accelerated scenario is the existence of a \textit{Dark Energy (DE)} component, characterized by an \textit{Equation of State (EoS)} with a value of $\omega_0=-1.018 \pm 0.057$ in the context of a flat Universe \cite{Planck2020}. The same observations have revealed a remarkable and somewhat perplexing fact: approximately 95-96\% of the content of the Universe exists in the form of two enigmatic components known as DE and \textit{Dark Matter (DM)}. In contrast, only a mere 4-5\% of the total composition is attributed to baryonic matter \cite{Peebles,Padmanabhan}. This discovery has highlighted the potential limitations of \textit{General Relativity (GR)}, which, while successful in explaining gravitational phenomena at the scale of the Solar System \cite{Will}, may prompt us to explore its applicability to gravitational phenomena on galactic and cosmological scales. The shortcomings of GR become evident when confronted with the two fundamental challenges that modern cosmology grapples with the DM and DE problems. 

The most prominent description of DE in the framework of GR is the cosmological constant $\Lambda$, which exhibits an EoS of $\omega_{\Lambda}=-1$ and can be associated with the vacuum quantum energy \cite{weinberg/1989}. While the cosmological constant aligns well with observational data, it is plagued by two major issues: the cosmic coincidence problem and the cosmological constant problem,
\begin{itemize}
    \item \textbf{Cosmic Coincidence Problem:} The cosmic coincidence problem questions why the energy densities of DM and DE in the Universe are comparable in magnitude at the present epoch. It appears unlikely that these two components, which evolve differently over time, would have similar magnitudes today without any underlying reason or mechanism \cite{dalal/2001}.
    \item \textbf{Cosmological Constant Problem:} The cosmological constant problem pertains to the extremely small but non-zero value of the cosmological constant, which is used to explain the accelerated expansion of the Universe. The problem lies in the significant discrepancy between the predicted and observed values of $\Lambda$. The theoretical estimates for $\Lambda$ are 120 orders of magnitude larger than the value suggested by observational data, resulting in a fine-tuning challenge \cite{Copeland}.
\end{itemize}

Recently, the cosmological landscape has been further complicated by the emergence of the so-called $4.9 \sigma$ 'Hubble tension' \cite{Valentino}. This tension arises from a significant discrepancy between the value of the Hubble constant, denoted as $H_0$ (expressed in $km/s/Mpc$), as measured by the Planck satellite ($67.4 \pm 0.5$) \cite{Planck2020} and by nearby standard candles such as SNeIa ($73.3 \pm 1.1$) \cite{Brout}. Despite the challenges posed by the cosmic coincidence problem and the fine-tuning issues associated with the cosmological constant, the $\Lambda$CDM model, where the DE density remains constant throughout the evolution of the Universe, stands as the most widely accepted cosmological model in contemporary astrophysics. To address these challenges, researchers have pursued two main approaches:
\begin{itemize}
    \item \textbf{Dynamical Approach:} This approach involves introducing dynamical models of DE and DM. Instead of assuming a cosmological constant, these models propose time-evolving scalar fields or other dynamical components to describe DE. Examples of dynamical DE models include quintessence \cite{Quin}, phantom \cite{Phan1,Phan2,Phan3}, k-essence \cite{ess}, and scalar-tensor theories \cite{scal}.
    \item \textbf{Modified Gravity:} The second approach explores modifications to the theory of gravity itself. Rather than introducing new components like DE or DM, this approach postulates alternative gravitational theories that could explain the observed cosmic acceleration without the need for additional exotic components. These modified gravity theories typically involve modifications to GR on cosmological scales. Examples include $f(R)$ gravity \cite{H.A.,H.K.,Odintsov1,Odintsov2,Odintsov3}, $f(T)$ gravity \cite{T1,T2,Odintsov4,Odintsov5}, and $f(Q)$ gravity \cite{Q0,Q1,Q2,Q3,Q4}.
\end{itemize}

Among the various approaches considered, one of the most promising and widely used methods is to employ the model-independent approach \cite{Shafieloo1,Shafieloo2}. Within the scientific literature, numerous physical arguments and motivations have emerged regarding this approach to investigating the dynamics of DE models. Model-independent approaches, as the name suggests, do not rely on assuming specific functional forms or parameterizations for DE or cosmological parameters. Instead, they allow for more freedom in exploring a broader range of possibilities and deviations from the standard $\Lambda$CDM model. In this study, we adopt a similar approach of cosmological parametrization, where we explicitly solve the field equations and explore the dynamics of the Universe during various phases of its evolution. Our aim is to describe specific phenomena, such as the cosmological phase transition from early inflation to deceleration, and subsequently from deceleration to late-time acceleration. To achieve this, we consider different parametrizations of cosmological parameters. These parametrizations involve model parameters that can be constrained using observational data. The majority of parametrizations in cosmology focus on characterizing the behavior of either the EoS parameter $\omega(z)$ \cite{Corasaniti}, or the deceleration parameter $q(z)$ \cite{Cunha}. Several well-known parametrizations include the Chevallier-Porrati-Linder (CPL) parametrization \cite{Chevallier}, the Jassal-Bagla-Padmanabhan (JBP) parametrization \cite{Jassal}, and the Barboza-Alcaniz (BA) parametrization \cite{Barboza}. These parametrizations primarily focus on characterizing the evolution of the EoS parameter $\omega(z)$. However, it is important to note that there are other geometric and physical parameters that can also be parameterized. A critical examination of this subject reveals that various other geometrical and physical parameters can also be subject to parametrization \cite{Mamon1,Koussour1,Campo}. Recently, Roy et al. \cite{Roy} investigated scalar field DE models using a general parametrization of the Hubble parameter. The study explores whether the observed cosmic acceleration can be described by quintessence or phantom scalar fields, offering insights into the nature of DE.

Motivated by the aforementioned discussions, this study explores two distinct parametrizations of the Hubble parameter $H(z)$: power-law and logarithmic corrections. The organization of this paper is as follows: In Sec. \ref{sec2}, we introduce the cosmological model within the context of a spatially flat FLRW Universe, considering the two parametrizations of the Hubble parameter. In Sec. \ref{sec3}, we delve into the observational data obtained from various sources, such as Cosmic Chronometers (CC), Baryonic Acoustic Oscillations (BAO), and the recently released Pantheon+ datasets. In addition, we discuss the Methodology employed to determine the model parameters. Sec. \ref{sec4} focuses on the investigation of the dynamics of essential cosmological parameters, including the deceleration parameter, energy density, pressure, and EoS parameter. Moreover, in Sec. \ref{sec5}, we use the $Om(z)$ diagnostic test, which provides a valuable tool for characterizing different cosmological models of DE. Lastly, in Sec. \ref{sec6}, we present a concise summary and our conclusions based on the findings from this study.

\section{Cosmological model}
\label{sec2}

In this article, we will focus on studying a specific type of Universe known as a spatially flat Friedmann-Lema\^{\i}tre-Robertson-Walker (FLRW) Universe, which describes an isotropic and homogeneous Universe. This type of Universe is characterized by a time-dependent scale factor denoted as $a(t)$. The FLRW metric is fundamental in enabling us to describe and understand the expansion of the Universe, forming a foundational framework in the field of cosmology. The metric for a spatially flat FLRW Universe is represented as \cite{Ryden}
\begin{equation}
ds^{2}=dt^{2}-a^{2}(t)[dr^{2}+r^{2}(d{\theta }^{2}+sin^{2}\theta d{\phi }%
^{2})],  \label{FLRW}
\end{equation}%
where $dt$ is the differential of time, and $dr$, $d\theta$, and $d\phi$ are differentials of spatial spherical coordinates. 

Moreover, the energy-momentum tensor for a perfect fluid, which we will consider in this context, is expressed in the following form
\begin{equation}
T_{\mu \nu }=\left( p+\rho \right) u_{\mu }u_{\nu }-pg_{\mu \nu }.
\label{EMT}
\end{equation}%

Here, the variables $u^{\mu}$, $\rho=\rho_{M}+\rho_{DE}$, $p=p_{M}+p_{DE}$, and $g_{\mu \nu }$ represent the four-velocity vector, the total energy density, the total pressure, and the metric tensor, respectively. The subscript "M" refers to matter, which includes both dark matter and baryonic matter, while the subscript "DE" represents dark energy. 

The Einstein field equations for GR are a set of equations that describe the relationship between the geometry of spacetime and the distribution of matter and energy within it. In their simplest form, the Einstein field equations are given by $G_{_{\mu \nu }}=\kappa T_{\mu \nu }$, where $\kappa =8\pi G=1$ and $G_{_{\mu \nu }}$ represents the Einstein tensor, which encapsulates the curvature of spacetime. By using Eqs. (\ref{FLRW}) and (\ref{EMT}), we can express the Einstein field equations for a spatially flat FLRW Universe as \cite{Myrzakulov}
\begin{equation}
3H^{2}=\rho  \label{F1}
\end{equation}%
\begin{equation}
2{\dot{H}}+3H^{2}=-p  \label{F2}
\end{equation}%
where $H=\frac{\dot{a}}{a}$ represents the current rate of expansion of the Universe.
Furthermore, Eqs. (\ref{F1}) and (\ref{F2}) are commonly referred to as the Friedmann equations. The first Friedmann equation establishes a connection between the expansion rate of the Universe and its energy density. It reveals how the energy content of the Universe influences the rate at which it is expanding. On the other hand, the second Friedmann equation relates the acceleration of the expansion rate to the pressure within the Universe. It describes how the presence of pressure, whether positive or negative, affects the change in the expansion rate over time.

To further understand the cosmic history and potential transitions to an accelerated period, we introduce the total equation of state (EoS) parameter $\omega$. This parameter is defined as the ratio of the total pressure to the total energy density:
\begin{equation}
\omega =\frac{p}{\rho }
\end{equation}

By using Eqs. (\ref{F1}) and (\ref{F2}), we can express the EoS parameter as
\begin{equation}
\omega =-\frac{2{\dot{H}}+3H^{2}}{3H^{2}}=-\frac{2}{3}\left(\frac{{\dot{H}}}{H^{2}}\right)-1.
\label{EoS}
\end{equation}

The EoS parameter $\omega$ provides valuable information about the nature of the dominant components driving the expansion of the Universe. It characterizes the behavior and properties of the cosmic fluids or fields that contribute to the total energy density. The EoS parameter allows us to categorize different types of matter and energy based on their pressure-to-density ratio. For example:
\begin{itemize}
    \item For matter with negligible pressure, such as non-relativistic matter (e.g., dark matter and baryonic matter), $\omega=0$.
    \item For radiation, which consists of relativistic particles like photons, $\omega=\frac{1}{3}$ due to the relationship between pressure and energy density.
    \item For a cosmological constant, which represents the energy associated with DE in $\Lambda$CDM model, $\omega_{\Lambda}=-1$, indicating a negative pressure that drives cosmic acceleration.
\end{itemize}

By using Eqs. (\ref{F1}) and (\ref{F2}), it is possible to derive the following expression \cite{Sahni1}
\begin{equation}
    \frac{\overset{..}{a}}{a}=-\frac{1}{6}\left( \rho +3p\right).
\end{equation}

Hence, according to the derived expression, the current model predicts acceleration ($\overset{..}{a}>0$) only when $\omega<-\frac{1}{3}$. In this accelerated phase of evolution, two distinct periods can be identified based on the value of $\omega$:  $-1/3 < \omega < -1$  corresponds to the quintessence phase, while  $\omega < -1$  signifies the onset of the phantom era.

To enable a more convenient comparison between theoretical results and cosmological observations, we introduce the redshift $z$ as an independent variable instead of the conventional time variable $t$. The redshift $z$ is defined by the following relation,
\begin{equation}
    1+z=\frac{1}{a(t)}.
\end{equation}

By imposing the condition that the present-day value of the scale factor is one ($a(0) = 1$), we can normalize the scale factor. Hence, we can express the derivatives with respect to time as derivatives with respect to the redshift using the following relation,
\begin{equation}
    \frac{d}{dt} =\frac{dz}{dt}\frac{d}{dz}=-\left( 1+z\right) H\left( z\right) \frac{d }{dz}.
\end{equation}

Moreover, the sign of the deceleration parameter $q$, provides information about whether the model undergoes decelerating or accelerating expansion. When $q > 0$, it indicates a deceleration in the expansion of the Universe. If $q = 0$, the expansion maintains a constant rate. On the other hand, when $-1 <q < 0$, it signifies accelerating expansion. Notably, when $q = -1$, the Universe experiences exponential expansion, known as de Sitter expansion. Furthermore, for $q < -1$, the Universe exhibits super-exponential expansion. The deceleration parameter can be defined as
\begin{equation}
    q(z)=-\frac{\overset{..}{a}}{aH^{2}}=\left( 1+z\right)\frac{1}{%
H\left( z\right) }\frac{dH\left( z\right) }{dz}-1.
\label{DP}
\end{equation}

As mentioned in the introduction, the standard $\Lambda$CDM model encounters two significant challenges: the cosmic coincidence problem and the cosmological constant problem. In the $\Lambda$CDM model, the Hubble parameter is expressed as
\begin{equation}
H(z) = H_0 \left[\Omega_{m0}(1+z)^3 + \Omega_{\Lambda} + \Omega_{r0}(1+z)^4\right]^\frac{1}{2},
\label{LCDM}
\end{equation}
where $H_0$ signifies the present-day expansion rate of the Universe, $\Omega_{m0}$ denotes the matter density parameter at the present epoch, $\Omega_{\Lambda}$ is the DE density parameter at the present epoch, and $\Omega_{r0}$ represents the radiation density parameter at the present epoch. While radiation played a significant role during earlier cosmic times, our analysis is centered on late-time evolution, where its contribution becomes negligible.

In our paper, we will employ a model-independent approach to the study of cosmological models and parameterization. The model-independent methodology offers the potential to reconstruct the entire cosmic history of the Universe and provides a framework for interpreting various cosmic phenomena. One of the key advantages of this approach is that it does not disrupt the background theory and offers a straightforward mathematical means to reconstruct the Universe's cosmic history. Furthermore, this strategy represents the most straightforward theoretical pathway to addressing several challenges within the Standard Model, such as the issue of initial singularities, the cosmological constant dilemma, and the persistent problem of decelerated expansion throughout cosmic history \cite{Cunha}. Several captivating models of DE and modified gravity have emerged, stemming from a range of parameterization schemes involving fundamental geometrical parameters, including the Hubble parameter, deceleration parameter, and jerk parameter \cite{Pacif1, Pacif2}. Hence, we investigate two different parametrizations of the Hubble parameter, specifically focusing on the correction terms associated with DE. These parametrizations, namely the power-law and logarithmic corrections, allow for deviations from the $\Lambda$CDM model at both low and high redshifts, 
\begin{eqnarray}
\label{H1} 
H_{1}(z)&=&H_0 \left[\Omega_{m0}(1+z)^3+B (1+z)^{\epsilon}\right]^\frac{1}{2}, \\
H_{2}(z)&=&H_0 \left[\Omega_{m0}(1+z)^3+B+\epsilon \log(1+z)\right]^\frac{1}{2},
   \label{H2} 
\end{eqnarray}
where $B$ and $\epsilon$ are free parameters introduced to account for corrections beyond the standard $\Lambda$CDM model. The parameter $\epsilon$ introduces power-law and logarithmic correction terms to the expansion rate, allowing for variations that evolve with redshift. While the precise physical origin of the parameter $\epsilon$ is not directly linked to a specific physical theory in our current study, its inclusion is motivated by the aim to explore potential deviations from the $\Lambda$CDM predictions. To achieve $H=H_0$ at $z=0$, it is necessary to impose the condition that $B=(1-\Omega_{m0})$. By satisfying this condition: $\epsilon=0$, the parameterizations given in Eqs. (\ref{LCDM}) can be reproduced. In addition, the term $B$ in the Hubble parameter models allows for deviations from the standard model specifically at low redshifts.

Now, using the Friedmann equations (\ref{F1}), (\ref{F2}) and Eqs. (\ref{H1}), (\ref{H2}), we can determine the expressions for the energy density and pressure in both models as
\begin{eqnarray}
\label{rho1} 
\rho_{1}(z)&=&3 H_{0}^{2} \left[\Omega_{m0}(1+z)^3+(1-\Omega_{m0}) (1+z)^{\epsilon}\right], \\
\rho_{2}(z)&=&3 H_{0}^{2} \left[\Omega_{m0}(1+z)^3+(1-\Omega_{m0})+\epsilon \log(1+z)\right],
   \label{rho2} 
\end{eqnarray}
and
\begin{eqnarray}
\label{p1} 
p_{1}(z)&=&-H_{0}^{2}\left[(\Omega_{m0}-1)(\epsilon -3) (1+z)^{\epsilon }\right], \\
p_{2}(z)&=&H_{0}^{2} \left[3 \Omega_{m0}-3 \epsilon  \log (1+z)+\epsilon -3\right],
   \label{p2} 
\end{eqnarray}

From Eqs. (\ref{EoS}), (\ref{H1}) and (\ref{H2}), we can analytically express the EoS parameter for both models as
\begin{eqnarray}
\label{EoS1} 
\omega_{1}(z)&=&\frac{(\Omega_{m0}-1) (\epsilon -3) (1+z)^{\epsilon }}{3 (\Omega_{m0}-1) (1+z)^{\epsilon }-3 \Omega_{m0} (1+z)^3}, \\
\omega_{2}(z)&=&\frac{3 \Omega_{m0}-3 \epsilon  \log (1+z)+\epsilon -3}{3 (\Omega_{m0} z (z (3+z)+3)+\epsilon  \log (1+z)+1)},
   \label{EoS2} 
\end{eqnarray}

Further, by using Eqs. (\ref{DP}), (\ref{H1}) and (\ref{H2}), we can calculate the deceleration parameter for both models as
\begin{eqnarray}
\label{DP1} 
q_{1}(z)&=&\frac{(\Omega_{m0}-1) (\epsilon -2) (1+z)^{\epsilon }-\Omega_{m0} (1+z)^3}{2 (\Omega_{m0}-1) (1+z)^{\epsilon }-2 \Omega_{m0} (1+z)^3}, \\
q_{2}(z)&=&-1+\frac{3 \Omega_{m0} (1+z)^3+\epsilon }{2 (\Omega_{m0} z (z (3+z)+3)+\epsilon  \log (1+z)+1)},
   \label{DP2} 
\end{eqnarray}

\section{Analysis of Observational Data and Methodology}
\label{sec3}

In the realm of observational cosmology, an integral aspect lies in constructing optimal cosmological models. To accomplish this, it becomes imperative to rigorously constrain the model parameters such as $\Omega_{m0}$, $\epsilon$ as well as the Hubble constant $H_0$, through the analysis of observational data. In this study, we employ a diverse range of observational datasets, encompassing Cosmic Chronometers (CC), Baryonic Acoustic Oscillations (BAO), and the latest Pantheon sample known as Pantheon+, derived from observations of Type Ia Supernovae (SNe). 

\subsection{Dataset \textit{Hz}: Cosmic Chronometers}
The Cosmic Chronometers (CC) method is a valuable technique used to measure the Hubble rate by studying the properties of the most ancient and passively evolving galaxies. These galaxies are carefully selected based on a small redshift interval, allowing for the implementation of the differential aging method. The Hubble rate $H$, defined within the FLRW metric is given by

$$H = -\frac{1}{1+z} \frac{dz}{dt}.$$

This relationship allows us to infer the rate of expansion of the Universe at different points in time. One of the key advantages of the CC method is its capacity to measure the Hubble parameter $H(z)$ without relying on specific cosmological assumptions. This attribute renders the CC method a valuable tool for testing and scrutinizing various cosmological models. Notably, R. Jimenez and A. Loeb \cite{jimlo} introduced a procedure that directly retrieves Hubble parameter data by computing the rate of redshift change, $dz/dt$, at a precise value of $z$. 

In this study, a comprehensive dataset comprising 31 data points has been meticulously compiled from a range of reputable surveys \cite{CC1,CC2,CC3,CC4,CC5,CC6,CC7,CC8}. These data points are derived using the CC method, covering a broad spectrum of redshift values ranging from 0.1 to 2, out of which 15 correlated data points within the range $0.179 < z < 1.965$ are derived from the measurements of \cite{CC4,CC6,CC7}. The covariance matrix linked with the CC method can be formulated as 

\begin{equation}
    Cov_{mn}=Cov_{mn}^{\text{S}}+Cov_{mn}^{\text{Y}}+Cov_{mn}^{\text{M}}+Cov_{mn}^{\text{SM}}.
\end{equation}

Here, the superscripts 'S', 'Y', 'M', and 'SM' represent the contributions to the covariance matrix arising from statistical errors, contamination by the young component, sensitivity to the chosen model, and variations in stellar metallicity, respectively. The contribution stemming from model-related covariance, $Cov_{mn}^{\text{M}}$, can be broken down further into segments originating from the star formation history (sfh), initial mass function (imf), stellar library (sl), and the considered stellar population synthesis (sps) model

\begin{equation}
    Cov_{mn}^{\text{M}}=Cov_{mn}^{\text{sfh}}+Cov_{mn}^{\text{imf}}+Cov_{mn}^{\text{sl}}+Cov_{mn}^{\text{sps}}.
\end{equation}

In this formulation, $Cov_{mn}^{\text{M}}$ signifies the covariance due to overall model uncertainties. Meanwhile, $Cov_{mn}^{\text{sfh}}$, $ Cov_{mn}^{\text{imf}}$, $ Cov_{mn}^{\text{sl}}$, and $Cov_{mn}^{\text{sps}}$ denote contributions originating from uncertainties in the star formation history, initial mass function, stellar library, and stellar population synthesis model, respectively. To perform an MCMC analysis, it is necessary to compute the chi-square function for correlated CC measurements, which is defined as

\begin{equation}
\chi^2_{cov}(\vartheta) = (\Delta H) Cov^{-1} (\Delta H)^{T},
\end{equation}
where $\Delta H_k = H_{th}(z_k,\vartheta)-H_{obs}(z_k)$.

Now, incorporating the utilization of the $\chi^2$ function for the remaining 16 non-correlated CC data points, we have

\begin{equation}
\chi^2_{noncov}(\vartheta)=\sum_{k=1}^{16}\left[\frac{(H_{th}(z_k,\vartheta)-H_{obs}(z_k))^2}{\sigma_H^2(z_k)}\right].
\end{equation}

Here, $H_{th}$ corresponds to the theoretical estimation of the Hubble parameter for a particular model, having the model parameters $\vartheta$. $H_{obs}$ signifies the observed values of the Hubble parameter and $\sigma_H$ denotes the associated error in the estimation.

Thus, the resultant $\chi^2$ function for $Hz$ data set is given by

\begin{equation}
    \chi^2_{Hz}=\chi^2_{cov}+\chi^2_{noncov}.
\end{equation}

\subsection{Dataset \textit{SNe}: Pantheon+}

The Pantheon+ analysis builds upon the original Pantheon analysis by incorporating an expanded dataset of supernova type Ia (SNe) that includes those with measured Cepheid distances to galaxies. This comprehensive dataset comprises 1701 light curves from 1550 SNe, spanning a redshift range $0.001\le z\le 2.2613$, sourced from 18 different studies \cite{pan1,pan2,pan3,pan4}. Out of the 1701 light curves in the dataset, 77 of them are associated with galaxies that contain Cepheids. Compared to the original Pantheon compilation by \cite{Scolnic}, the Pantheon+ compilation introduces significant enhancements. Primarily, it features an expanded sample size, particularly for SNe at redshifts below 0.01. Additionally, notable improvements have been made in addressing systematic uncertainties associated with redshifts, intrinsic scatter models, photometric calibration, and peculiar velocities of SNe. It is significant that due to specific selection criteria, not all SNe from the original Pantheon compilation are included in the enhanced Pantheon+ compilation. 

Determining the Hubble constant ($H_0$)  with the highest precision and shift in its extracted value involves a method that centers around the utilization of SNe observations. These observations are calibrated through the use of Cepheid variable stars present in galaxies hosting both SNe and Cepheid variables. To establish the calibration of Cepheids, geometric methods like parallax are employed, both within our own Milky Way and in nearby anchor galaxies. This distance ladder method provides a direct measurement of the $H_0$. Pantheon+ offers an additional advantage by providing the ability to constrain the $H_0$ along with model parameters. This capacity arises from the incorporation of the distance moduli of SNe located in Cepheid host galaxies. These distance moduli are directly derived from the distance ladder analysis performed by the SH0ES team \cite{pan1}. 

Additionally, Pantheon+ encompasses the consideration of covariance between these SNe and those situated within the Hubble flow. In contrast to the initial Pantheon sample, which encountered limitations in estimating $H_0$ due to the degeneracy between $H_0$ and the absolute magnitude $M$ of SNe, the improved results of the Pantheon+ method overcome this challenge. By combining both the apparent magnitude $m_B$ and the distance modulus $\mu^{cd}_k$ derived from Cepheids associated with SNe in Cepheid host galaxies, the absolute magnitude $M = m_{Bk} - \mu^{cd}_k$ can be independently determined. This decoupling of the degeneracy between $M$ and $H_0$ provides the means to independently assess the value of $H_0$ through the Pantheon+ dataset.

In order to obtain the best fits for the free parameters, it is necessary to optimize the $\chi^2$ function, which is expressed as 

\begin{equation}
    \chi^2_{SNe}= \Delta\mu^T (C_{Sys+Stat}^{-1})\Delta\mu.
\end{equation}

Here, $C_{Sys+Stat}$ represents the covariance matrix of the Pantheon$^+$ dataset, encompassing both systematic and statistical uncertainties. 

$\Delta\mu$ denotes the distance residual and is defined by

\begin{equation}
    \Delta\mu_k=\mu_k-\mu_{th}(z_k).
\end{equation}

In the above equation, $\mu_k$ signifies the distance modulus of the $k^{th}$ SNe. It is important to note that $\mu_k$ is calculated as $\mu_k=m_{Bk}-M$, where $m_{Bk}$ represents the apparent magnitude of the $k^{th}$ SNe and $M$ denotes the fiducial magnitude of an SNe. 

The theoretical distance modulus $\mu_{th}$ is determined using the exppression:

\begin{equation}\label{Eq:mu}
    \mu^{th}(z,\vartheta)=5\log_{10}\left(\frac{d_L(z,\vartheta)}{1\text{ Mpc}} \right)+25,
\end{equation}

where $d_L$ denotes the model-based luminosity distance in Mpc, given by

\begin{equation}
    d_L(z,\vartheta)=\frac{c(1+z)}{H_0}\int_0^z \frac{d\zeta}{E(\zeta)}.
\end{equation}

Here, $c$ represents the speed of light and $E(z)=\frac{H(z)}{H_0}$.

Further, the distance residual is represented by

\begin{equation}
    \Delta\Bar{\mu}=\begin{cases}
            \mu_k-\mu_k^{cd}, & \text{if $k$ is in Cepheid hosts}\\
            \mu_k-\mu_{th}(z_k), & \text{otherwise}
         \end{cases}.
\end{equation}

Here, $\mu_k^{cd}$ refers to the Cepheid host-galaxy distance released by SH0ES. While calculating the covariance matrix for the Cepheid host- galaxy, it can be combined with the covariance matrix for SNe. The combined covariance matrix, denoted as $C^{SNe}_{Sys+Stat}+C^{cd}_{Sys+Stat}$, encompasses both statistical and systematic uncertainties from the Pantheon+ dataset. Thus, the $\chi^2$ function for the combined covariance matrix employed to constrain cosmological models in the analysis is given by

\begin{equation}\label{eq:chiSNe}
    \chi^2_{SNe+}= \Delta\Bar{\mu} (C^{SNe}_{Sys+Stat}+C^{cd}_{Sys+Stat})^{-1}\Delta\Bar{\mu}^T.
\end{equation}

\subsection{Dataset \textit{BAO}: Baryonic Acoustic Oscillations}
Baryonic Acoustic Oscillations (BAO) represent fluctuations in the density of baryonic matter in the Universe, resulting from acoustic density waves in the primordial plasma during the early stages of the Universe. These oscillations provide valuable insights as they can be utilized to extract important cosmological parameters related to DE. By analyzing the BAO peaks in the matter power spectrum, the Hubble distance $D_H(z)$ and the angular diameter distance $D_A(z)$ can be determined. The sound horizon $r$, associated with the BAO peaks, allows for the computation of the angular separation $\delta_\theta=r/(1+z)D_A(z)$ and the redshift separation $\delta_z=r/D_H(z)$ at a particular redshift $z$. These quantities play a crucial role in characterizing the spatial distribution of matter and constraining cosmological models. By carefully selecting appropriate values of $r$ and effectively constraining the parameters governing the ratios $D_H(z)/r$ and $D_A(z)/r$, one can accurately estimate the Hubble parameter $H(z)$ at different redshifts. Through several BAO data surveys \cite{BAO1,BAO2,BAO3,BAO4,BAO5,BAO6,BAO7,BAO8,BAO9,BAO10,BAO11,BAO12} (see Table 4 of \cite{DMNaik}) one can calibrate the Hubble value. Various surveys, such as Delubac et al. (BOSS) \cite{BAO1}, Blake et al. (WiggleZ) \cite{BAO2}, Chuang et al. (SDSS III) \cite{BAO3}, Ribera et al. (BOSS) \cite{BAO4}, and others, provide measured values of the angular distance function $D_A$ or the Hubble distance function $D_H$. These values are employed to deduce the Hubble parameter. Notably, the calculation of the Hubble parameter in some original articles involves specific selections of $r$. For instance, in \cite{BAO1}, $r$ is chosen as $147.4 ~Mpc$, while in \cite{BAO3}, it follows the value from D. J. Eisenstein (1998) \cite{soundhorizon}; \cite{BAO4} on the other hand, adopts a $r$ value of $147.49 ~Mpc$. In these studies, the Hubble parameter is evaluated based on the respective $r$ choices. 

In the present analysis, we consider the BAO measurements from \cite{cao} (ref Table 1) and \cite{alam} (ref Table 3) to constrain the model parameters.

 To find the $\chi^2$ function for BAO, we use the following relation

\begin{equation}
    \chi^2_{BAO}(\vartheta)=\sum_{k=1}^{12}\left[\frac{(H_{th}(z_i,\vartheta)-H_{obs}(z_k))^2}{\sigma_H^2(z_k)}\right],
\end{equation}

For joint analysis of Hz, SNe, and BAO dataset, the $\chi^2_{joint}$ function is defined as
\begin{equation}
    \chi^2_{joint}=\chi^2_{Hz}+\chi^2_{SNe}+\chi^2_{BAO}.
\end{equation}

\subsection{Observational Results}

The best-fit parameters are determined by minimizing the $\chi^2$ function, which is related to the likelihood through $\mathcal{L} \propto \exp \left( -\frac{\chi^2}{2} \right)$. By utilizing the Markov Chain Monte Carlo (MCMC) sampling technique with Python's {\tt emcee} library, we obtain numerical constraints on the model parameters. The results are presented as contour plots, illustrating confidence levels up to $2$-$\sigma$ based on the likelihood analysis.

\begin{figure}[!]
     \centering
     \includegraphics[width=\linewidth]{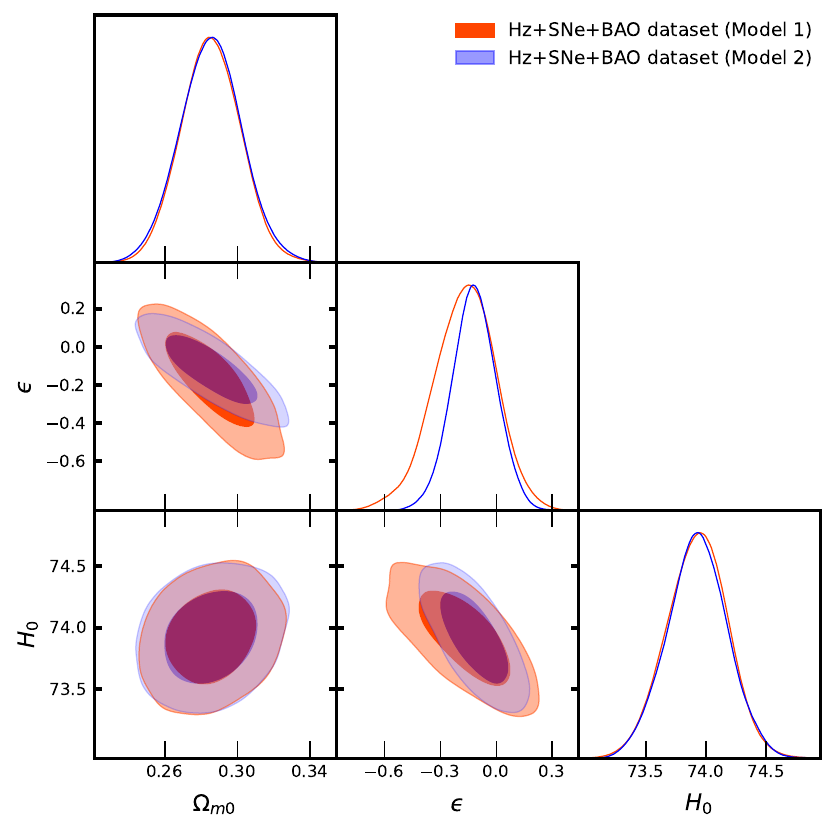}
     \caption{2D likelihood contour for the joint analysis of Hz, SNe, and BAO datasets up to 2$\sigma$ error. The Orange contour represents Model 1, and the blue contour represents Model 2.}\label{fig:combine}
\end{figure}

\begin{figure*}[!]
     \centering
     \subfloat[\label{fig:hz1}]{\includegraphics[width=0.48\linewidth]{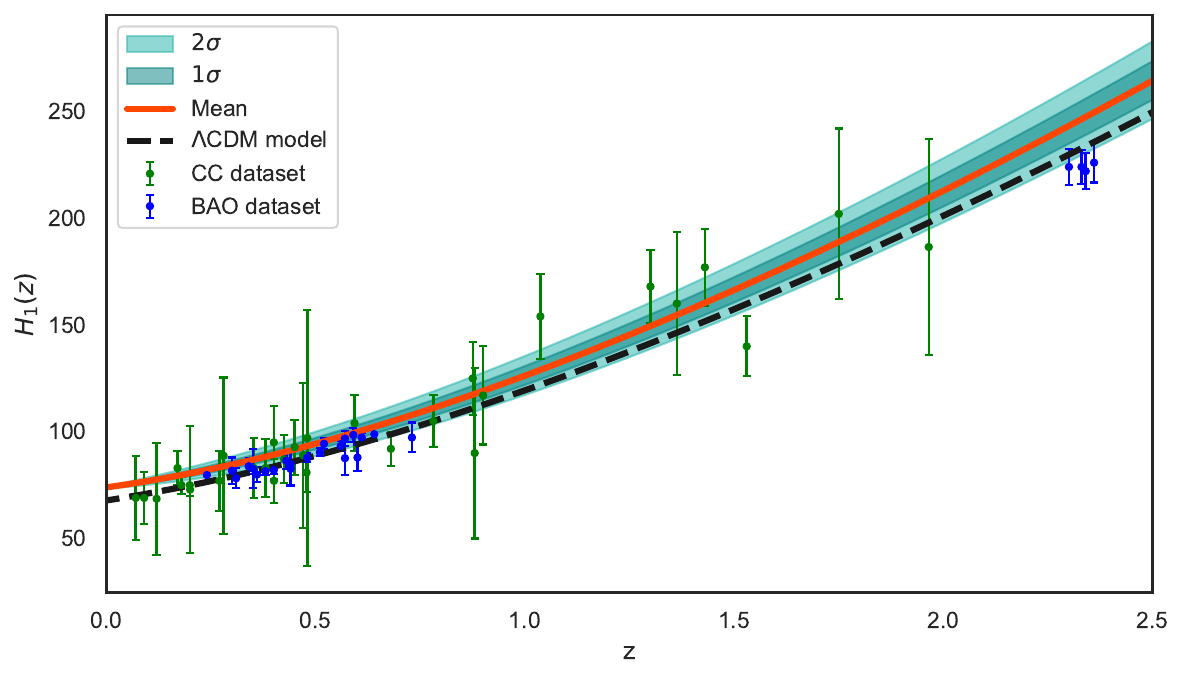}}
     \subfloat[\label{fig:hz2}]{\includegraphics[width=0.48\linewidth]{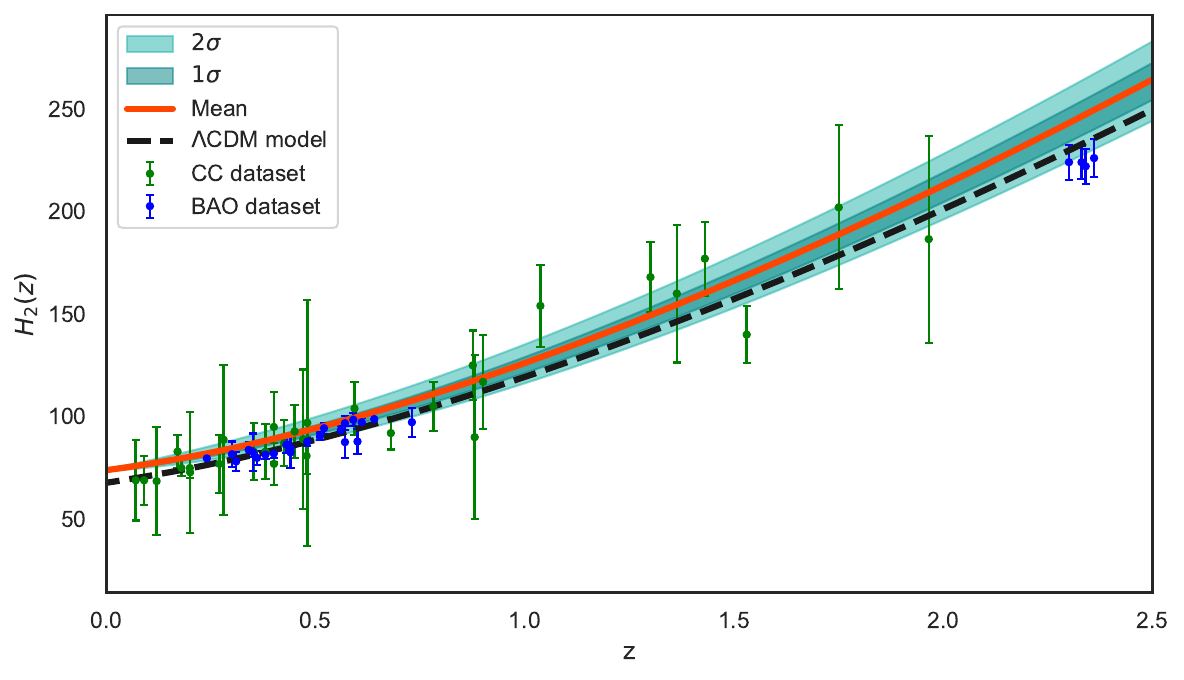}}
     \caption{Plot showing the behavior of the Hubble parameters based on the values obtained from the joint analysis of Hz, SNe, and BAO datasets. The error bands represent the $1\sigma$ and $2\sigma$ regions, the red curve corresponds to the mean values of the parameters and the black dotted line corresponds to $\Lambda$CDM model with $\Omega_\Lambda=0.7$, $\Omega_m=0.3$ and $ H_0= 67.8$.}\label{fig:hz}
\end{figure*}

\begin{figure*}[!]
     \centering
     \subfloat[\label{fig:mz1}]{\includegraphics[width=0.48\linewidth]{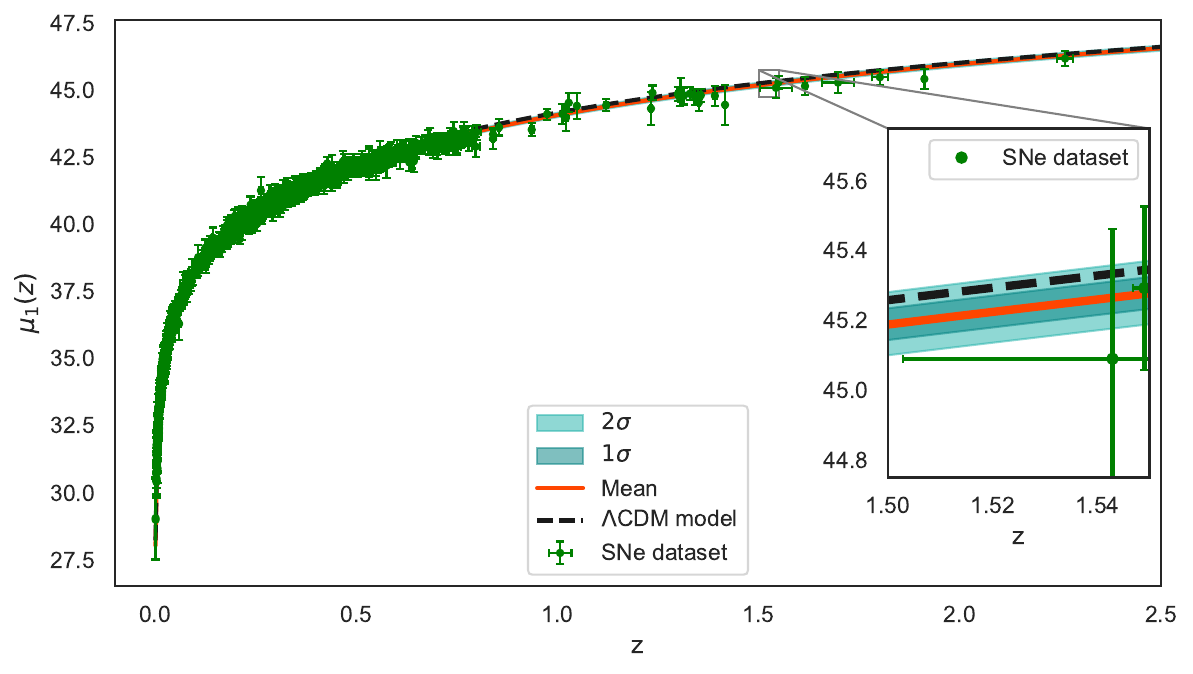}}
     \subfloat[\label{fig:mz2}]{\includegraphics[width=0.48\linewidth]{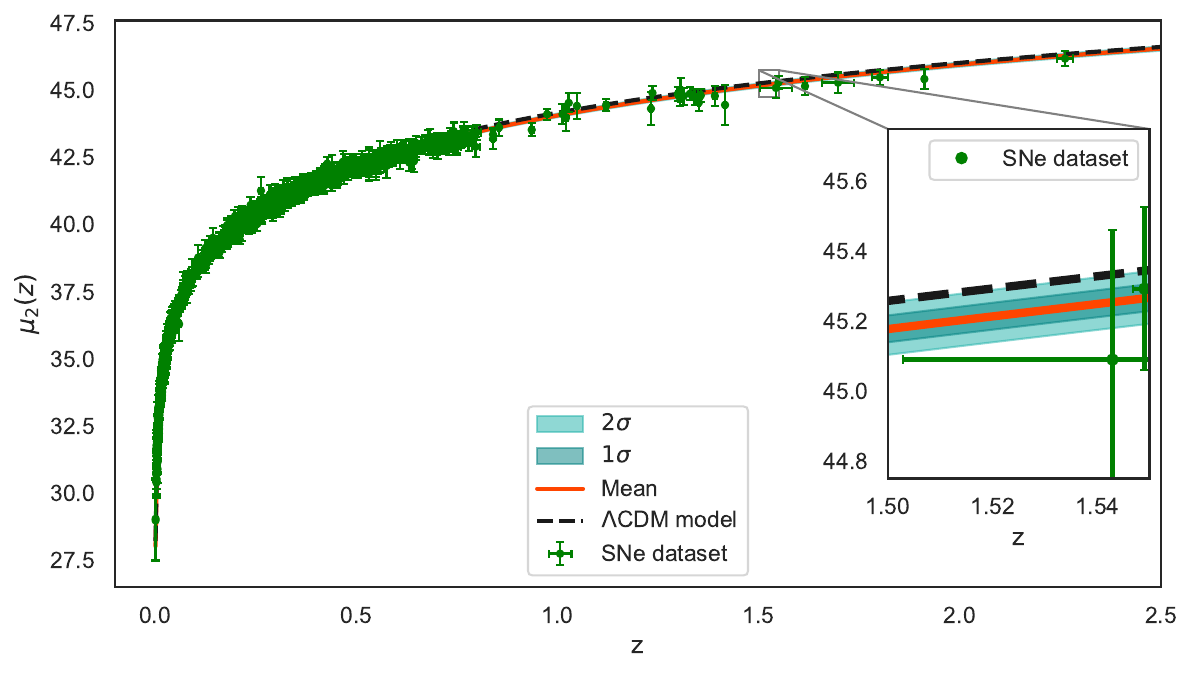}}
     \caption{Plot showing the behavior of distance modulus functions based on the values obtained from the joint analysis of Hz, SNe, and BAO datasets. The error bands represent the $1\sigma$ and $2\sigma$ regions, the red curve corresponds to the mean values of the parameters and the black dotted line corresponds to $\Lambda$CDM model with $\Omega_\Lambda=0.7$, $\Omega_m=0.3$ and $ H_0= 67.8$.}\label{fig:mz}
\end{figure*}

The analysis yields the following mean values for the model parameters: $\Omega_{m0}=0.285\pm 0.016$, $\epsilon=-0.17^{+0.17}_{-0.15}$, and $H_0=73.92\pm0.25$ for power-law parametrization (Model 1) and $\Omega_{m0}=0.285\pm 0.017$, $\epsilon=-0.12\pm0.12$, and $H_0=73.93\pm0.25$ for logarithmic parametrization (Model 2). Figure~\ref{fig:combine} showcases the 2D likelihood contours up to $2\sigma$ errors corresponding to these parameter values. A summary of the MCMC results can be found in Table~\ref{tab:tab_1}. It is worth noting that our model demonstrates consistency with all the utilized datasets in the analysis. Figure~\ref{fig:hz} showcases 1$\sigma$ and 2$\sigma$ bounds of the theoretical curves for the parameterized Hubble functions along with the observed $H(z)$ data. The error bars representing the observed distance modulus of the 1701 SNe data points, as well as the corresponding best-fit theoretical curves for distance modulus functions, are presented in Figure~\ref{fig:mz}. These figures depict the comparison between our parameterized models and the standard $\Lambda$CDM model. 

\begin{table*}[t]
    \caption{Summary of the results from the MCMC analysis for the parameters $\Omega_{m0}$, $\epsilon$, and $H_0$.}
    \label{tab:tab_1}
	\centering
	\begin{tabularx}{\linewidth}{>{\centering\arraybackslash}X >{\centering\arraybackslash}X >{\centering\arraybackslash}X >{\centering\arraybackslash}X}
        \hline
        \hline
		Model &  $\Omega_{m0}$ & $\epsilon$ & $H_0$\\ 
		\hline
            Model 1 & $0.285\pm 0.016$ & $-0.17^{+0.17}_{-0.15}$ & $73.92\pm0.25$\\
            Model 2 & $0.285\pm 0.017$ & $-0.12\pm0.12$ & $73.93\pm0.25$ \\
            \hline
            \hline
	\end{tabularx}
    \end{table*}

\subsection{Bayesian Model Comparison}

To identify the optimal model among the two proposed parametrizations, we evaluate them using two widely accepted information criteria for model selection, the Akaike Information Criterion (AIC) \cite{akaike} and the Bayesian Information Criterion (BIC) \cite{akaike1}. Both AIC and BIC are commonly employed in statistics and data science for this purpose, each imposing distinct penalties for additional parameters, thereby potentially leading to disparate model selection outcomes. AIC, based on the likelihood function of the data and the model's parameter count, serves as a means to balance model complexity and goodness of fit. The AIC is mathematically expressed as
\begin{equation}
    AIC = 2p - 2\ln \mathcal{L}_{max},
\end{equation}
where $\mathcal{L}_{max}$ represents the maximum likelihood and $p$ denotes the number of parameters in the model employed for MCMC. The model with the lowest AIC, denoted as AIC$^{Best}$, is considered the optimal choice. To gauge the level of support for the $n$th model, we compute the difference between AIC$^n$ and AIC$^{Best}$, denoted as $\Delta$ AIC$^n$. If $\Delta$ AIC$^n$ is less than 2, it implies that the $n$th model is nearly as effective as the best model. However, a $\Delta$ AIC$^n$ between 4 and 7 suggests considerably weaker support for the $n$th model. When $\Delta$ AIC$^n$ exceeds 10, it signals that the $n$th model is improbable to be the optimal choice and is therefore recommended for exclusion from consideration.

Similar to AIC, BIC also penalizes models with a higher number of parameters while favoring those that exhibit better data fit. However, BIC imposes a more substantial penalty on additional parameters compared to AIC. The BIC is computed as follows
\begin{equation}
    BIC = p\ln N - 2\ln \mathcal{L}_{max},
\end{equation}
where $N$ denotes the number of data points used in the analysis. The identification of the optimal model involves selecting the one with the minimum BIC value, denoted as BIC$^{Best}$. To assess the relative support for the $n$th model, we calculate $\Delta$BIC$^n$ as the difference between the BIC value of the $n$th model and the BIC value of the best model (BIC$^{Best}$).

With $M$ models under consideration, the magnitude of $\Delta$BIC serves as evidence against the $n$th model being the optimal choice. A $\Delta$BIC$^n$ less than 2 suggests very weak evidence favoring the $n$th model over the best model. Values between 2 and 6 indicate positive evidence against the $n$th model, while a range of 6 to 10 suggests strong evidence against it. A $\Delta$BIC$^n$ greater than 10 signifies very strong evidence, implying that the $n$th model is improbable as the optimal choice.

The $\chi_{\text{min}}^2$ values for model 1 and model 2 are $1741.5290$ and $1741.8759$ respectively. Based on this, we choose model 2 as our reference model. We now compare our model 1 with the reference model. The value $\Delta$AIC $\Delta$BIC for model 1 is $0.3469$. Further, it is crucial to examine the $\Delta$AIC and $\Delta$BIC values for the $\Lambda$CDM model. Using the equations for the computation of AIC and BIC, we have $\Delta$AIC and $\Delta$BIC for the $\Lambda$CDM model \eqref{LCDM} as $1.3883$, meaning our model fits the data similar to that of $\Lambda$CDM model.

\section{Dynamics of Cosmological Parameters}
\label{sec4}

Cosmological parameters play a fundamental role in shaping our understanding of the Universe and its evolution. These parameters are essential quantities used to describe the fundamental properties and dynamics of the cosmos. They encompass various aspects, including the geometry of space, the composition of matter and energy, the expansion rate, and the state of cosmic expansion. The role of cosmological parameters extends beyond theoretical modeling; they are also vital for interpreting observational data. By comparing theoretical predictions based on specific parameter values with observed cosmic phenomena, cosmologists can test and refine cosmological models. This interplay between theory and observation allows us to constrain the values of cosmological parameters. 

In this section, we delve into an examination of the cosmological parameters that encompass energy density, pressure, the EoS parameter, and the deceleration parameter. All of these cosmological parameters are plotted with respect to the redshift ($z$) of the model parameters, constrained through a joint analysis of Hz, SNe, and BAO datasets for both the power-law and logarithmic models. The plotted figures also include their corresponding $1\sigma$ and $2\sigma$ error bands.

\begin{figure*}[!]
     \centering
     \subfloat[\label{fig:rho1}]{\includegraphics[width=0.43\linewidth]{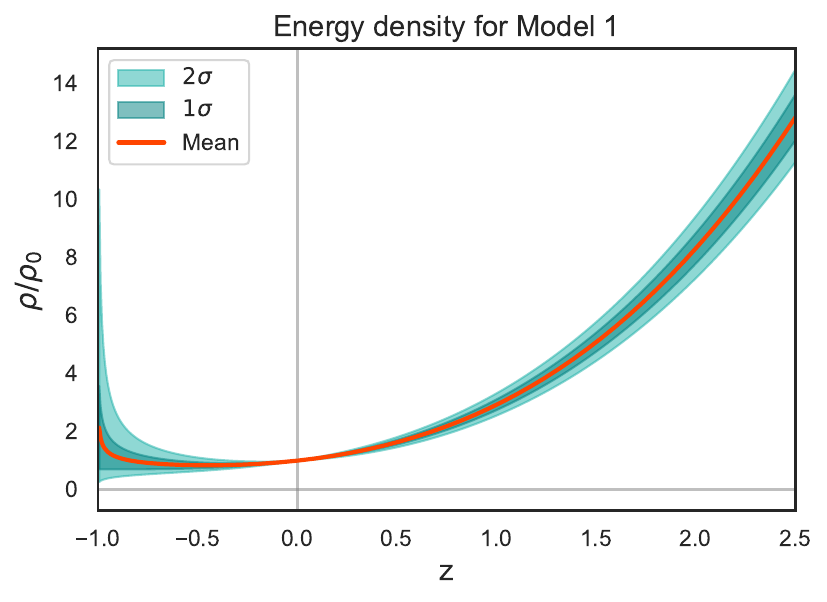}}
     \subfloat[\label{fig:rho2}]{\includegraphics[width=0.43\linewidth]{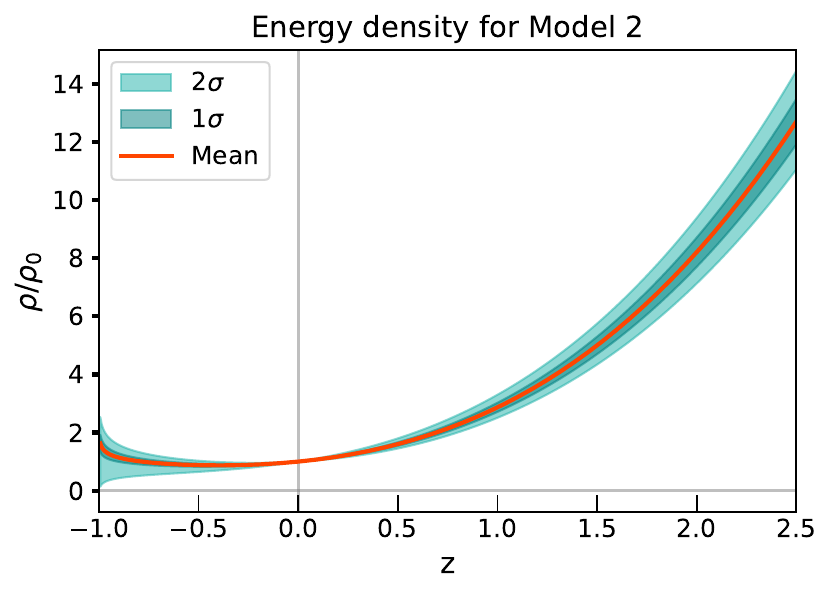}}
     \caption{Plot showing the behavior of energy densities based on the values obtained from the joint analysis of Hz, SNe, and BAO datasets. The error bands represent the $1\sigma$ and $2\sigma$ regions, and the red curve corresponds to the mean values of the parameters.}\label{fig:rho}
\end{figure*}

\begin{figure*}[!]
     \centering
     \subfloat[\label{fig:p1}]{\includegraphics[width=0.43\linewidth]{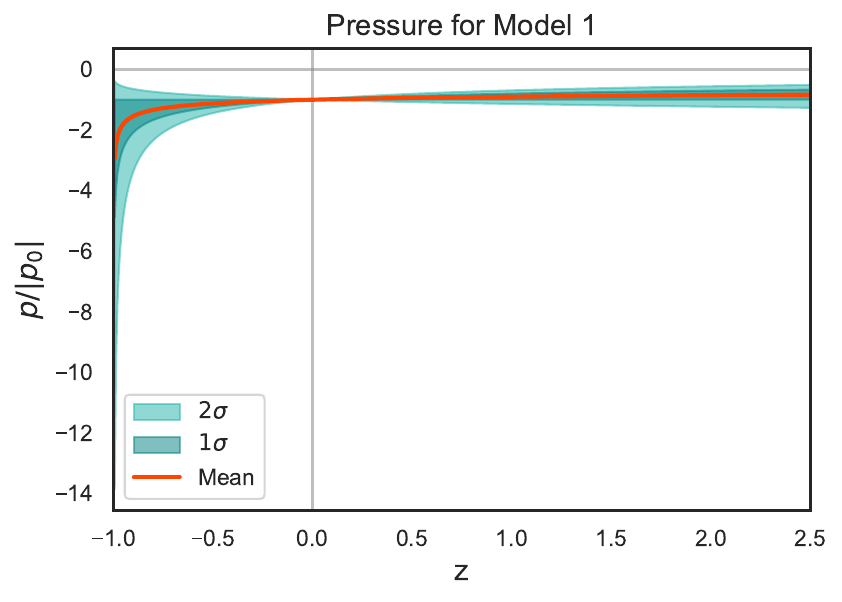}}
     \subfloat[\label{fig:p2}]{\includegraphics[width=0.43\linewidth]{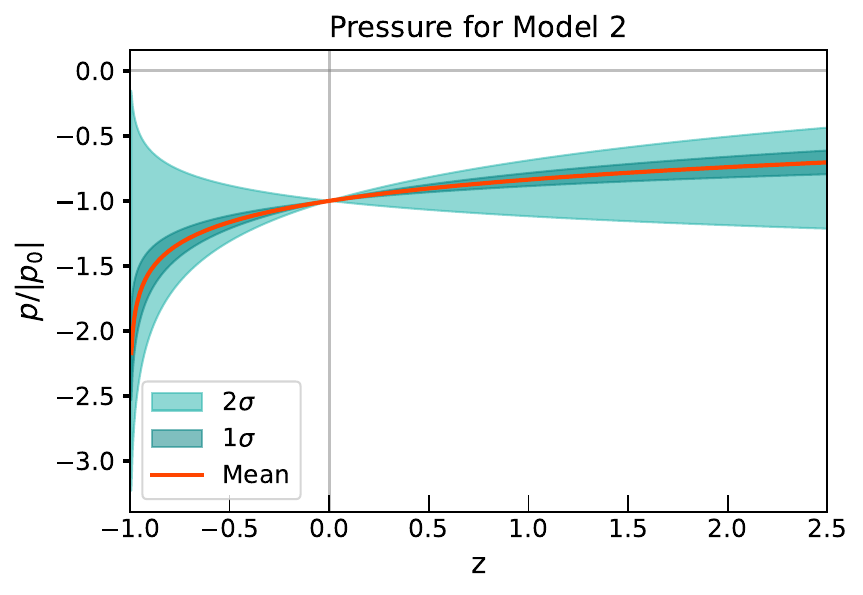}}
     \caption{Plot showing the behavior of pressures based on the values obtained from the joint analysis of Hz, SNe, and BAO datasets. The error bands represent the $1\sigma$ and $2\sigma$ regions, and the red curve corresponds to the mean values of the parameters.}\label{fig:p}
\end{figure*}

Figs. \ref{fig:rho1} and \ref{fig:rho2} provide the expected behavior of energy densities in both models. As anticipated, the energy densities demonstrate a positive trend and gradually decrease as the Universe expands. This behavior is consistent with our understanding of the present and far future of the Universe, where energy densities naturally diminish over time. The plots in Figs. \ref{fig:p1} and \ref{fig:p2}, on the other hand, reveal a negative behavior of the pressure for both models. This negative behavior signifies the occurrence of late-time cosmic acceleration in the Universe, aligning with the current understanding that the expansion of the Universe is accelerating. The negative pressure indicated in these plots supports the presence of a form of DE that drives the observed cosmic acceleration. Together, these findings from Figs. \ref{fig:rho1}, \ref{fig:rho2}, \ref{fig:p1} and \ref{fig:p2} provide valuable evidence for the expected behavior of energy densities and the late-time cosmic acceleration of the Universe in both models.

\begin{figure*}[!]
     \centering
     \subfloat[\label{fig:eos1}]{\includegraphics[width=0.43\linewidth]{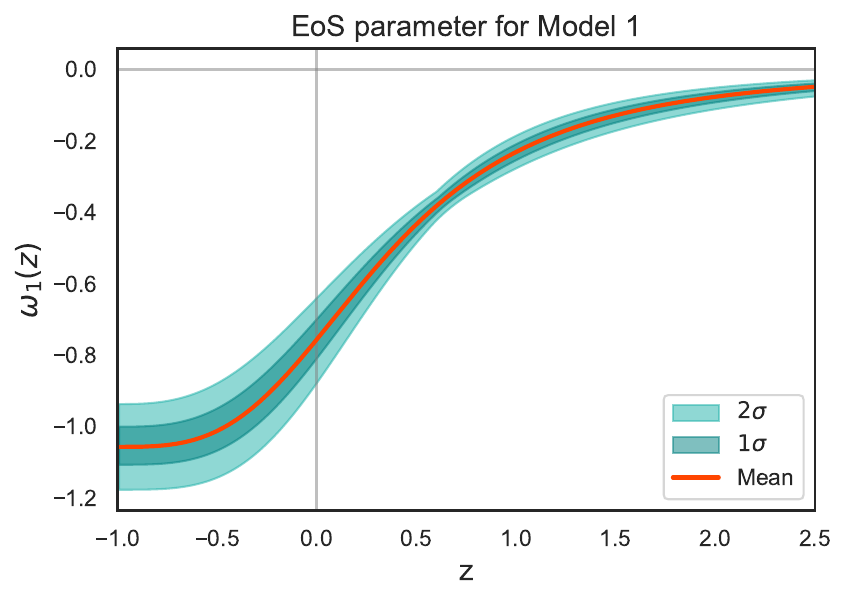}}
     \subfloat[\label{fig:eos2}]{\includegraphics[width=0.43\linewidth]{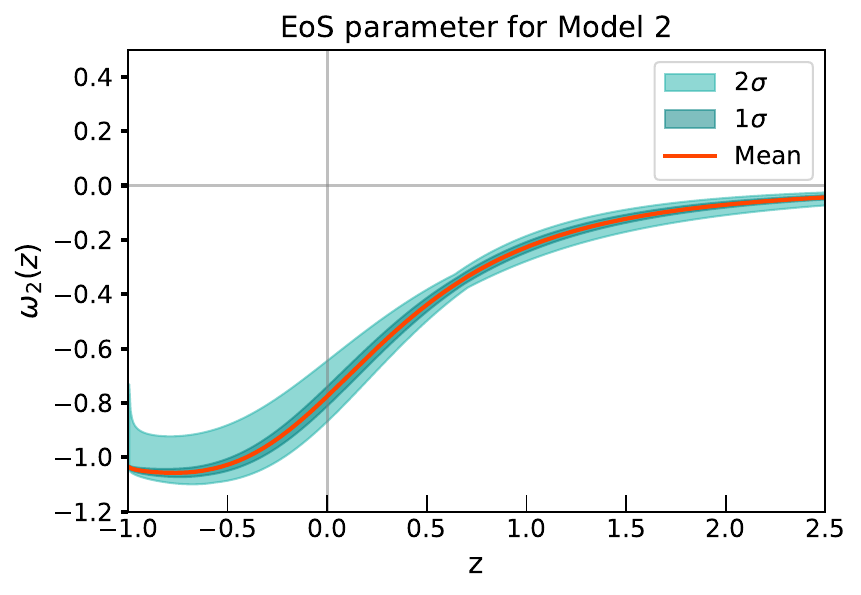}}
     \caption{Plot showing the behavior of EoS parameters based on the values obtained from the joint analysis of Hz, SNe, and BAO datasets. The error bands represent the $1\sigma$ and $2\sigma$ regions, and the red curve corresponds to the mean values of the parameters.}\label{fig:eos}
\end{figure*}

The EoS parameter serves as a valuable tool for characterizing the different phases of the expanding Universe. In our analysis, EoS parameters have been derived for both Model 1 and Model 2. These parameters provide insights into the nature of the dominant components driving the evolution of the Universe. Fig. \ref{fig:eos1} illustrates the behavior of the EoS parameter for Model 1. It is noteworthy that the EoS parameter approaches the behavior of the $\Lambda$CDM model for lower values of $z$. This alignment with the $\Lambda$CDM model suggests that Model 1 exhibits similar characteristics to the standard cosmological model in the future Universe. As the redshift increases, the EoS parameter indicates a matter-dominated era of the Universe, consistent with the expected behavior during the matter-dominated epoch. In the present epoch, the EoS parameter exhibits quintessence-like behavior, implying the presence of a slowly evolving scalar field driving the current accelerated expansion of the Universe \cite{Quin}. Moving on to Fig. \ref{fig:eos2}, we observe similar behavior in the graph of the EoS parameter for Model 2. The EoS parameter follows a comparable pattern to that discussed for Model 1, further supporting the consistency of their respective behaviors. Through the process of fitting the model to observational data, we have determined the present value of the EoS parameter. The constrained values of the model parameters yield an EoS parameter of $\omega_{0}=-0.756_{-0.053}^{+0.057}$ for Model 1 and $\omega_{0}=-0.755 _{-0.057}^{+0.015}$ for Model 2 \cite{Hernandez,Gong,Zhang}.

\begin{figure*}[!]
     \centering
     \subfloat[\label{fig:q1}]{\includegraphics[width=0.43\linewidth]{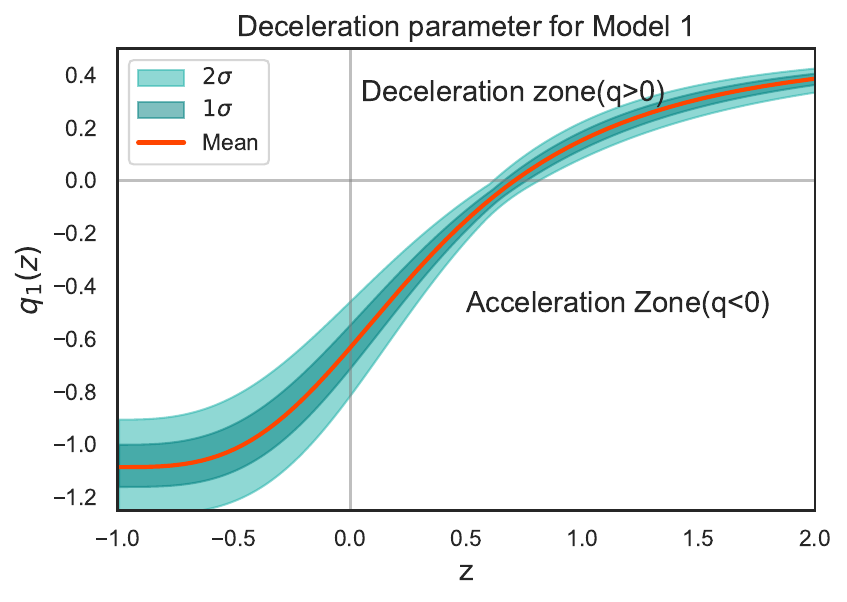}}
     \subfloat[\label{fig:q2}]{\includegraphics[width=0.43\linewidth]{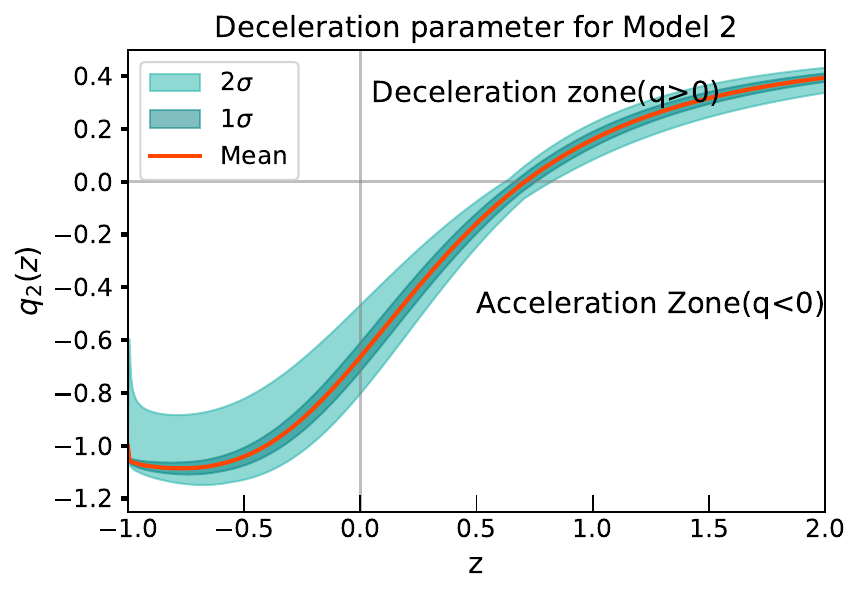}}
     \caption{Plot showing the behavior of the deceleration parameters based on the values obtained from the joint analysis of Hz, SNe, and BAO datasets. The error bands represent the $1\sigma$ and $2\sigma$ regions, and the red curve corresponds to the mean values of the parameters.}\label{fig:q}
\end{figure*}

The dynamics of the deceleration parameter for Model 1 exhibit a positive behavior for higher values of $z$, as depicted in Fig. \ref{fig:q1}. As the value of $z$ decreases, we observe a transition to the negative behavior of the deceleration parameter. This transition signifies that Model 1 undergoes a decelerated phase during the early times of the Universe, followed by an accelerated phase as $z$ decreases. Similar behavior can be observed for Model 2, as illustrated in Fig. \ref{fig:q2}, where the deceleration parameter exhibits the same trend. This consistent behavior of the deceleration parameter in both models indicates a decelerated phase in the early Universe and a transition to an accelerated phase as redshift decreases. The value of the transition redshift $z_{tr}$ for both models exhibits fluctuations within the range of 0.3 to 1.0, as indicated by recent observational data  \cite{Mehrabi}. Further, the present value of the deceleration parameter has been evaluated to have a value of $q_{0}=-0.633 _{-0.080}^{+0.085}
$ for Model 1 and $q_{0}=-0.633 _{-0.086}^{+0.022}
$ for Model 2 \cite{Garza,Cunha,Camarena}.

\section{$Om(z)$ diagnostics}
\label{sec5}

The $Om(z)$ diagnostic serves as a valuable tool for categorizing various cosmological models of DE \cite{Omz}. This diagnostic is particularly appealing due to its simplicity, as it relies solely on the first-order derivative of the cosmic scale factor. In the case of a spatially flat Universe, the $Om(z)$ diagnostic can be expressed as 
\begin{equation}
Om\left( z\right) =\frac{\left( \frac{H\left( z\right) }{H_{0}}\right) ^{2}-1%
}{\left( 1+z\right) ^{3}-1}.
\end{equation}

The behavior of the $Om(z)$ function can be interpreted based on the slope of its curve. A negative slope indicates quintessence-type behavior, where the DE component behaves similarly to a slowly evolving scalar field. Conversely, a positive slope corresponds to phantom behavior, suggesting the presence of an extremely negative pressure that drives an exponentially accelerating expansion. In contrast, a constant value of $Om(z)$ reflects the behavior predicted by the $\Lambda$CDM model, where DE is represented by a cosmological constant. In this case, the energy density of DE remains constant throughout the evolution of the Universe.

Using Eqs. (\ref{H1}) and (\ref{H2}), we can mathematically express the $Om(z)$ diagnostic for both models as
\begin{eqnarray}
\label{Om1} 
Om_{1}(z)&=&\frac{-\left((\Omega_{m0}-1) (1+z)^{\epsilon }\right)+\Omega_{m0} (1+z)^3-1}{(1+z)^3-1}, \\
Om_{2}(z)&=&\Omega_{m0}+\frac{\epsilon  \log (1+z)}{z (z (3+z)+3)}.
   \label{Om2} 
\end{eqnarray}

\begin{figure*}[!]
     \centering
     \subfloat[\label{fig:om1}]{\includegraphics[width=0.43\linewidth]{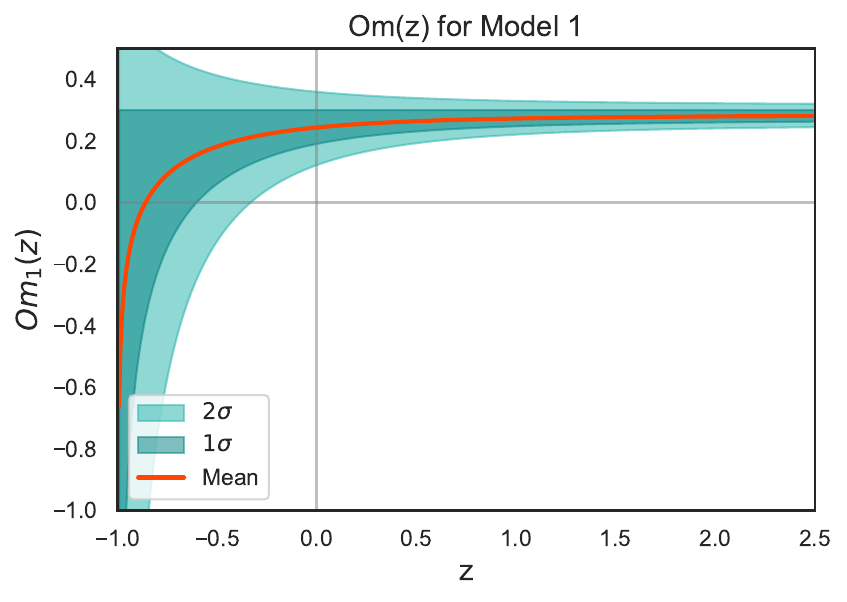}}
     \subfloat[\label{fig:om2}]{\includegraphics[width=0.43\linewidth]{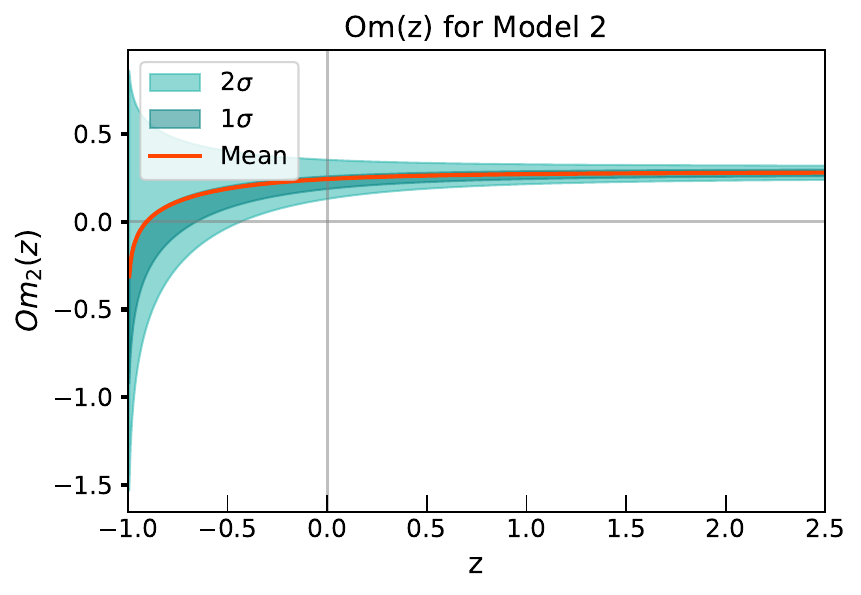}}
     \caption{Plot showing the behavior of $Om(z)$ based on the values obtained from the joint analysis of Hz, SNe, and BAO datasets. The error bands represent the $1\sigma$ and $2\sigma$ regions, and the red curve corresponds to the mean values of the parameters.}\label{fig:om}
\end{figure*}

Using the same set of model parameters obtained from the joint analysis of Hz, SNe, and BAO datasets, we further examine the behavior of the $Om(z)$ diagnostic. In Fig. \ref{fig:om1}, the $Om(z)$ diagnostic is displayed for Model 1, while Fig. \ref{fig:om2} presents the $Om(z)$ diagnostic for Model 2. Both plots exhibit a positive slope, indicating that both Model 1 and Model 2 behave as phantom. So, the positive slope of the $Om(z)$ diagnostic suggests that the dominant component in the Universe, described by these models, possesses properties similar to phantom. 

We now examine the CPL model, a widely recognized parametrization introduced in references \cite{cpl1, cpl2}. The model is mathematically defined by the following equation 

\begin{equation}\label{eq:cpl}
    \omega(z)_{cpl} = b + a\frac{z}{1+z}.
\end{equation}

To explore its implications, we perform a statistical MCMC analysis using observational datasets. The value of $\chi^2_{min}$ so obtained is $1738.0245$. The primary reason for the choice of a well-known CPL model is its ability to explain quintessence as well as a phantom state \cite{cpl1, cpl2}. Now, using the Bayesian model comparison technique discussed earlier, $\Delta$AIC and $\Delta$BIC assuming the same reference model turn out to be $3.5045$ which is a greater value than $2$, implying the evidence against the model is positive.

\section{Conclusions}
\label{sec6}

In this study, we have investigated the dynamics of the Universe's expansion and the behavior of DE using two parametrizations of the Hubble parameter: power-law and logarithmic corrections. By employing a model-independent approach \cite{Shafieloo1,Shafieloo2} and confronting the cosmological model with observational data from 31 points of CC samples, 1701 points of Pantheon+, and 12 points of BAO samples, we have constrained the model's parameters $\Omega_{m0}$, $\epsilon$, and $H_0$ using MCMC analysis (see Tab. \ref{tab:tab_1}). The best-fit values of the model parameters and the corresponding $1\sigma$ and $2\sigma$ confidence regions are depicted in Fig. \ref{fig:combine}. Our analysis reveals that the cosmological model with power-law and logarithmic corrections presents a compelling fit to the recent observational data (Figs. \ref{fig:hz1}, \ref{fig:hz2}, \ref{fig:mz1}, and \ref{fig:mz2}). It effectively describes the observed cosmic acceleration scenario and offers a correction to the standard $\Lambda$CDM model.

Moreover, the study of essential cosmological parameters, including the energy density, pressure, deceleration parameter, and EoS parameter, offers deeper insights into the behavior of DE. The behavior of cosmological parameters for both the power-law and logarithmic corrections models exhibits intriguing characteristics. In the power-law model, we observed notable deviations from the standard $\Lambda$CDM model. The energy density and pressure exhibit distinct positive and negative behaviors, respectively (Figs. \ref{fig:rho1}, \ref{fig:p1}). The deceleration parameter indicates transitions between different cosmic phases (Fig. \ref{fig:q1}). Moreover, the EoS parameter demonstrates intriguing dynamics, exhibiting quintessence-like behavior (Fig. \ref{fig:eos2}). We observed similar behavior for the logarithmic model (Figs. \ref{fig:rho2}, \ref{fig:p2}, \ref{fig:eos2}, and \ref{fig:q2}). A notable observation is that the models showcase contrasting evolutionary paths for the cosmological parameters in the future, while displaying similar behavior in the past. As we examine the trajectories of the parameters over time (or redshift), it becomes evident that the models deviate in their predictions for the future evolution of the Universe.

Finally, we applied the $Om(z)$ diagnostic test to both the power-law and logarithmic corrected models to categorize and classify their behavior (Figs. \ref{fig:om1} and \ref{fig:om2}). For both models, the $Om(z)$ diagnostic displayed a distinct positive slope, indicating phantom-like behavior, which is consistent with the observed accelerated expansion of the Universe.

\textbf{Data availability} All data used in this study are cited in the references and were obtained from publicly available sources.

\end{document}